# TRANSPORT IN MAGNETIC NANOPARTICLE SUPER-LATTICES : COULOMB BLOCKADE, MAGNETORESISTANCE, HYSTERESIS AND MAGNETIC FIELD INDUCED SWITCHING

Reasmey P. Tan[1], Julian Carrey[1,*], Céline Desvaux[2], Jérémie Grisolia[1], Philippe Renaud[2], Bruno Chaudret[3], Marc Respaud[1,*]

[1] Laboratoire de Physique et Chimie des Nano-Objets, INSA, 135, av. de Rangueil, 31077 Toulouse cedex, France
[2] Semiconductor Products Sector, Freescale, le Mirail B.P. 1029, 31023 Toulouse Cedex, France
[3] Laboratoire de Chimie de Coordination-CNRS, 205 rte de Narbonne, 31077 Toulouse cedex 4, France

## Abstract :

We report on magnetotransport measurements on millimetric super-lattices of Co-Fe nanoparticles surrounded by an organic layer. At low temperature, the transition between the Coulomb blockade and the conductive regime becomes abrupt and hysteretic. The transition between both regime can be induced by a magnetic field, leading to a novel mechanism of magnetoresistance. Between 1.8 and 10 K, high-field magnetoresistance attributed to magnetic disorder at the surface of the particles is also observed. Below 1.8 K, this magnetoresistance abruptly collapses and a low-field magnetoresistance is observed.

Spin-polarized transport processes, 72.25.–b
Electrical conductivity of disordered solids, 72.80.Ng

## TEXT:

Three-dimensional (3D) super-lattices of nanoparticles (NPs) are artificial solids whose building blocks are NPs surrounded by organic or inorganic thin layers. Recent progress in synthesis methods has enabled the fabrication of large super-lattices with building blocks that are metallic [1], semiconductor [2], magnetic [3] or even binary NPs [4]. In all cases, such a long range organisation is only possible using highly monodisperse particles [5].

When such super-lattices are composed of ferromagnetic (FM) NPs surrounded by an insulating layer, they display magneto-resistive properties which may have applications in high-density non-volatile memories or magnetic sensors [6]. Two different types of magnetoresistance (MR) have been reported in such systems [7-9]. First, the electron transfer rate between two spin-polarized neighbouring particles depends on the angle between their magnetization, leading to tunnel MR [7]. Second, the presence of magnetic disorder at the surface of the particles can increase the resistance of the particle array. The lifting of this disorder by a magnetic field leads to a negative high-field MR that we will refer to as "spin-disorder" MR [8,9]. In chemically-prepared NPs, the insulating layer is generally organic. Only a few molecules have been experimentally shown to be efficient spin-conservative tunnel barrier [2,7,10,11].

With respect to their electronic properties, super-lattices of metallic NPs are well described by arrays of localized electrons interacting through long-range Coulomb forces [12,13]. Notably, when the thermal energy is lower than the Coulomb interaction energy, electrical conduction is strongly reduced below a threshold voltage $V_T$, a phenomenon known as Coulomb blockade [12,14-17]. In the case of an array of particles, $V_T$ depends on the size of the particles and on their separation, as well as on the size of the array [12,14]. Experimental realizations of arrays of localized electrons such as arrays of NPs always present a certain amount of disorder, due to structural inhomogeneities or to offset electrical charges in the matrix. When the thermal energy is much lower than the Coulomb interaction energy, theoretical studies have shown that a glassy state emerges from the interplay between the electrostatic interactions and this disorder [18-22]. By analogy with the spin glass, this state has been called "Coulomb glass" or "electron glass". Experimental evidences of a Coulomb glass formation has been reported in a few systems in the last decade [20,22,23,24] but never in arrays of NPs, although they constitute the prototypical realization of an array of localized electrons.

In this article, we report on the magnetotransport measurements performed on 3D super-lattices of FM FeCo NPs surrounded by organic ligands. Transport properties are typical of Coulomb blockade in arrays of NPs. However below a critical temperature, unusual features appear in the current-voltage characteristics such as abrupt and hysteretic transitions between the Coulomb blockade and the conductive regime. The transitions between the two regimes can be triggered by a magnetic field, leading to a novel mechanism of MR. Beside, the super-lattices display more conventional MR properties : (*i*) between 1.8 and 10 K, a high-field MR attributed to spin-disorder MR is observed; (*ii*) this high-field MR abruptly collapses below 1.8 K and a low-field MR is observed.

The samples under study are composed of metallic FM monodisperse Cobalt-Iron (Co-Fe) NPs synthesized using organometallic chemistry and organized inside very large 3D super-lattices with fcc packing (see inset of Fig. 1a) [3]. The super-lattices have the shape of small chips or needles, the length (thickness) of which ranges between 1 and 2 mm (0.5 to 1 mm). The particles are 15 nm in diameter, separated by a 2 nm organic barrier composed of a mixture of hexadecylamine and long chain carboxylic acids. The magnetization at saturation is $M_S = 163$ $A.m^2.kg_{FeCo}^{-1}$ at 2 K, below the expected bulk FeCo magnetization (240 $A.m^2.kg_{FeCo}^{-1}$). The lack of magnetization is a consequence of an imperfect alloying between Fe and Co [3]. A coercive field of 0.03 T and a saturation field of 0.8 T are measured at 2 K on a single superlattice when the magnetic field is applied perpendicularly to the needles.

The super-lattices were hand-connected using Au wires and silver painting in a glove box. The time of transfer between the glove box and the cryostat was kept to a few tens of seconds to avoid oxidation. In the following experiments, the magnetic field was applied perpendicularly to the current and to the easy axis of the superlattice, with a sweep rate of 0.009 T/s. As-synthesized super-lattices display a room temperature resistance ranging from 8 kΩ to 5 GΩ. The origin of this dispersion is still unclear. All the samples displayed high-field spin-disorder MR between 1.8 and 10 K and low-field MR below 1.8 K. However, the novel mechanism of MR was only observed when the room temperature resistance of the samples was below 300 kΩ. All the results presented here arise from a single super-lattice displaying a low resistance (8 kΩ at 300 K).

Resistance-temperature ($R(T)$) and current-voltage ($I(V)$) characteristics of the samples present features typical of Coulomb blockade. Resistance measured at a voltage of 10 mV increases exponentially with decreasing temperature (see Fig. 1a). In an array of NPs in the Coulomb blockade regime, the low-bias resistance $R$ follows $R = R_O \exp[(T_O/T)^\nu]$, where $R_0$ is the high-temperature resistance and $T_0$ the activation temperature for charge transport. Plotting $\ln[-(\partial \ln R)/(\partial \ln T)]$ as a function of $\ln(T)$ allows a precise determination of $\nu = 0.48$ (± 0.01). It is very close to the value $\nu = 1/2$, generally measured in granular metals or in arrays of NPs with structural and/or charge disorder [25]. Fig. 1b shows that $\ln(R)$ varies linearly with $T^{-1/2}$ between 150 and 1.8 K. From the slope of the curve, the activation energy $T_0 = 24$ (± 2) K is extracted. A theoretical estimate of $T_0$ can be done using $k_B T_0 = e^2/2C$ and $C = 2N\pi\varepsilon_O \varepsilon_r \, r \ln(1 + 2r/s)$, where $r$ and $s$ are the particle radius and separating distance respectively, $\varepsilon_r$ the dielectric constant of the organic layer, and $N$ the number of nearest neighbours [26]. Using the values of $r = 7.5$ (±0.5) nm, $s = 2$ (±0.5) nm deduced from TEM observations, $N = 12$ and taking $\varepsilon_r = 2.2$ [7], we find $T_0 = 39$ (±7) K, in the same range as the experimental value.

$I(V)$ characteristics measured at various temperatures are presented in Fig. 1c. The Coulomb gap progressively opens when the temperature is lowered. When the temperature is further lowered below a critical temperature $T_{cr} = 1.8$ K, the transition between the Coulomb blockade regime below $V_T$ and the conductive one above $V_T$ becomes abrupt and hysteretic. The hysteresis area, the $V_T$ value and the amplitude of the jump increase strongly as the temperature is slightly lowered.

When a magnetic field is applied to the sample at 1.5 K, the hysteresis loop of the $I(V)$ characteristic slightly shifts toward the left, evidencing a small decrease of $V_T$ (see Fig 1d). This suggests that a magnetic field could directly induce the transition between the two regimes at the condition that the voltage applied to the sample is close but below $V_T$. Fig. 2 shows the experimental evidence of such transitions. At 1.5 K and without applied magnetic field, $V_T$ equals 18.45 V. When the applied voltage is 250 mV below this value, a 5 T magnetic field is necessary to induce the abrupt transition from the initial resistive state to the conductive state (see Fig. 2a). This transition is irreversible, i.e. the sample remains in the conductive state when the field is swept back to zero (not shown). The critical field of 5 T and the irreversible character of the transition are both in agreement with the $I(V)$ characteristics shown in Fig. 1d. When the applied voltage is very close to $V_T$ (about 20 mV), a smaller magnetic field of 70 mT induces the transition (see Fig. 2a). In both experiments, the two states of conduction in the Fig. 2a correspond to the two possible states of resistance at a given voltage in the $I(V)$ characteristic (see Fig. 2b). In Fig. 2c and 2d, we show similar experiments performed at 1.8 K, a temperature for which the hysteresis area and the threshold voltage are smaller. In this case, the magnetic field-induced transition is less abrupt and reversible. The noise observed on this curve is not due to our experimental setup but is intrinsic to the sample. Indeed, at this temperature, the electrical noise increases dramatically when the sample is close to the transition, a phenomenon which was also observed in some $I(V)$ characteristics. We emphasize that this magnetic field induced transition between two states is not observed i) when the applied voltage is too far from $V_T$ ii) at higher temperature above $T_{cr}$.

Other MR properties of the samples are now detailed. In the following, the MR is defined as $(R_{high} - R_{min}) / R_{min}$, where $R_{high}$ ($R_{min}$) is the high (low) resistance of the $R(H)$ characteristic. Between 1.8 and 10 K, a high-field MR is observed (see Fig. 3b for a typical $R(H)$ curve). The peaks appearing at $\mu_0 H = 0.5$ T and the butterfly shape of the $R(H)$ characteristic are both due to dynamical effects: when the resistance of the sample is measured as a function of time after varying the magnetic field, a lag of about 90 seconds between the variation of magnetic field and the variation of resistance is observed. The amplitude of the high-field MR depends on the applied voltage (see the squares in Fig. 3c). The complete MR(*V*) dependence, extracted from two *I*(*V*) characteristics measured at $\mu_0 H = 0$ T and $\mu_0 H = 2.6$ T is shown in Fig. 3c. The amplitude of this high-field MR varies from sample to sample and reaches regularly values well above 100 % and up to 3000 % at low voltage in one of them. Fig. 3a displays the evolution of this high-field MR as a function of temperature. Two well-defined regimes are observed: the MR amplitude regularly increases when decreasing temperature from 10 K down to 1.8 K and abruptly drops to a much lower level below 1.8 K.

Below 1.8 K, a vanishing high-field MR is still observed, the amplitude of which is only about 2 %/T at 2 V, with still a strong voltage dependence. Superimposed on this background a small-amplitude low-field MR is observed. The low-field *R*(*H*) shape depends on whether the sample is in the resistive regime or in the conductive one, as shown in Fig. 3d. The low-field MR seems to be composed of two different contributions. A first one appears as a peak centred around zero magnetic field, and the second one as a typical inverse tunnel MR characteristic, with a minimum at a field around 0.1 T and a saturation field around 1 T. This last contribution almost disappears when the sample is in the conductive regime. As a consequence, only the peak centred around zero is visible in the *R*(*H*) characteristic measured at 20 V.

We now discuss the origin of these various types of MR, starting with the high-field MR (see Fig. 3a, 3b and 3c). Similar MR has been previously observed in magnetic thin films or arrays of FM NPs [8,9,27-32]. In some cases [8,32], these effects have been related to the spin polarisation of the material. In our case, it is unlikely that such a large amplitude (up to 3000 %) could be related to the spin polarisation of the material [8]. Instead, we attribute this behaviour to spin-disorder MR [9,27-31,33]. The collapse of the spin-disorder MR below 1.8 K is thought to be a consequence of the freezing of the surface magnetic moments, which may undergo a paramagnetic-FM or a paramagnetic-glass transition below 1.8 K.

The low-field MR observed below 1.8 K (see Fig. 3d) is complex and is currently not fully understood. The contribution to this MR only visible at low-voltage could be due to spin-dependant tunnelling between particles since the saturation field of this contribution matches with the one observed in magnetic measurements. However the field corresponding to its minimum (100 mT) is higher than the one observed in magnetic measurements (30 mT), which may come from the higher sweep rate used during transport measurements. The inverse sign for the tunnel MR may be due to tunnelling via impurity states [34, 35]. The peak around zero visible both at low and high voltage is not explained yet.

The novel mechanism of MR shown in Fig. 2 is directly related to the hysteresis observed in the *I*(*V*) characteristics. It is independent of the magnetic state of the particles since it is observed above or below the saturation field of the NPs. Such transitions had never been reported in arrays of NPs in the Coulomb blockade regime. Unexplained transitions in the *I*(*V*) characteristics of CdSe nanorods arrays have been observed but they were not correlated to Coulomb blockade [36]. Hysteretic switching is also a well-known property of some molecules. However, these phenomena occur at a much higher voltage than the voltage drop between two NPs in our case. We think that the most likely hypothesis to explain our experimental results is the formation of a glassy state or a charge-ordered state inside the sample below $T_{cr}$. Indeed, history-dependant phenomena and avalanches are predicted and observed in strongly interacting systems with disorder such as Coulomb glasses [20, 22-24] and spin glasses [37]. Coulomb glass has been theoretically predicted to develop between $0.045 T_0$ and $0.095 T_0$, depending on the degree of disorder of the array [22]. The opening of the hysteretic behaviour starts in our case at $T_{cr} \approx 0.07 T_0$, using the $T_0$ value deduced from the $R(T)$ curve.

Finally, we discuss the fact that this type of MR was only observed in the more conductive samples. We speculate that the low resistive samples may have a thinner tunnel barrier between the particles so they would be more strongly coupled, making easier the apparition of collective effects. Another explanation is that the hysteresis and jumps would occur at voltage higher than the maximum voltage we can apply in our experimental setup (200 V). Indeed, more resistive samples display a wider Coulomb gap than the less resistive ones.

To conclude, these transport measurements reveals the richness and complexity of the conductivity mechanisms in hybrid artificial solids containing FM NPs surrounded by molecules. The confirmation of the origin of the abrupt and hysteretic transitions in the conductivity deserves further experimental studies, especially by changing the nature and the size of the NPs.

## Acknowledgements :

We thank M. Bibes, P. Seneor, R. Matana, J.-M. Georges, F. Nguyen Vandau and A. Fert for fruitful discussions about these results.

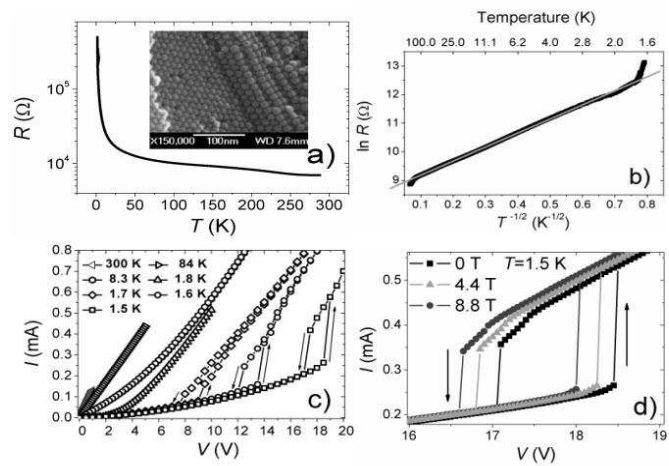

FIG 1. (a) Resistance $R$ as a function of temperature measured at a constant voltage $V = 10$ mV. (inset) Scanning electron micrograph of a typical super-lattice (b) $\ln(R)$ plotted as a function of $T^{-1/2}$ (c) $I(V)$ characteristics at various temperatures. Arrows indicate the sweep direction. (d) Influence of the magnetic field on the $I(V)$ characteristics measured at 1.5 K. $H = 0$ T (right curve), 4.4 T (middle curve) and 8.8 T (left curve).

Fig. 1

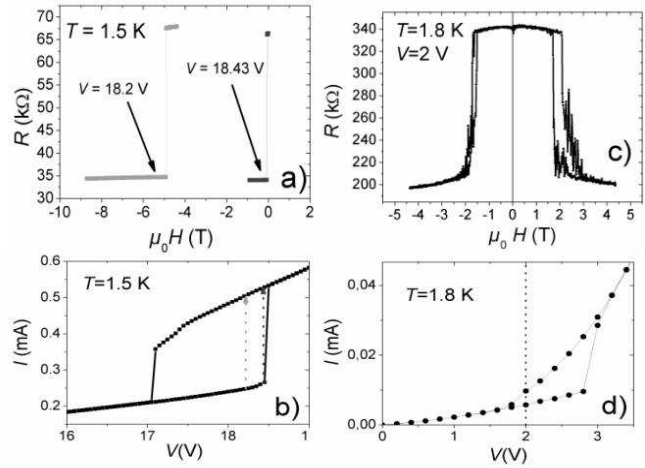

FIG. 2. (a) MR characteristic measured at 1.5 K with $V = 18.43$ V and $V = 18.2$ V. (b) Enlarged view of the $I(V)$ characteristic at 1.5 K. The two arrows show the transition induced by the magnetic field in (a). (c) MR characteristic measured at 1.8 K with $V = 2$ V (d) Enlarged view of the $I(V)$ characteristic at 1.8 K. The dotted line shows the voltage at which the MR in (c) has been measured.

# Fig. 2

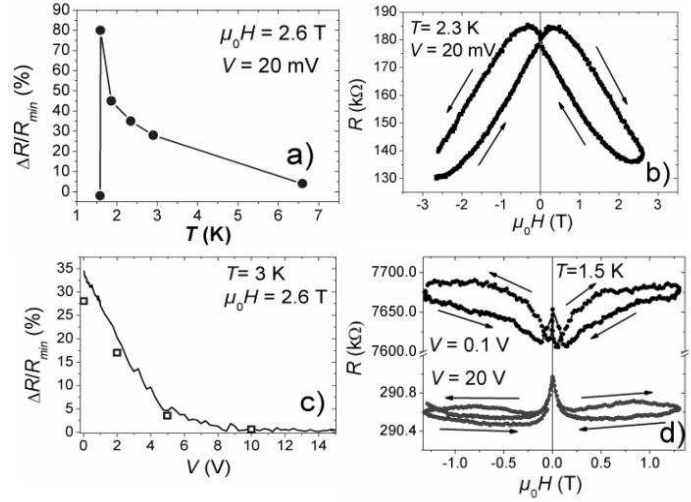

FIG. 3. (a) High-field MR amplitude as a function of temperature, measured at 2.6 T and $V = 20$ mV. (b) $R(H)$ characteristic for $V = 20$ mV and $T = 2.34$ K. (c) MR amplitude as a function of the applied voltage at $T = 3$ K deduced from two I(V) characteristics at 0 and 2.6 T. Squares shows the MR amplitude deduced from individual measurements at various voltages.
(d) Low-field MR at $T = 1.5$ K for $V = 0.1$ V (top curve) and $V = 20$ V (down curve). Note the break in the left axis.

Fig. 3